\documentclass[aps,prl,twocolumn,showpacs,groupedaddress]{revtex4}

\usepackage{graphicx}

\newcommand{\etal}{{\it et~al.}}
\newcommand{\eg}{{\it e.g.}}
\newcommand{\dwave}{$d_{x^2-y^2}$}
\newcommand{\icb}{$I_c(H_a)$}
\newcommand{\tm}{{$T=4.2$~K}}
\newcommand{\nccoy}{Nd$_{2-x}$Ce$_{x}$CuO$_{4-y}$}

\newcommand{\nccoop}{Nd$_{1.85}$Ce$_{0.15}$CuO$_4$}
\newcommand{\nccoopy}{Nd$_{1.85}$Ce$_{0.15}$CuO$_{4-y}$}
\newcommand{\nccoov}{Nd$_{1.835}$Ce$_{0.165}$CuO$_4$}
\newcommand{\nccoovy}{Nd$_{1.835}$Ce$_{0.165}$CuO$_{4-y}$}

\newcommand{\pccoy}{Pr$_{2-x}$Ce$_{x}$CuO$_{4-y}$}
\newcommand{\lccoy}{La$_{2-x}$Ce$_{x}$CuO$_{4-y}$}
\newcommand{\ybco}{YBa$_2$Cu$_3$O$_7$}

\begin{document}


\title{Phase-sensitive order parameter symmetry test experiments utilizing \nccoy/Nb zigzag junctions}

\author{Ariando}

\author{D. Darminto}

\altaffiliation[Permanent address: ]{Department of Physics,
Faculty of Mathematics and Sciences, Sepuluh November Institute of
Technology, Surabaya 60111, Indonesia}

\author{H. -J. H. Smilde}

\altaffiliation[Present address: ]{CEA-DRT-LETI - CEA/GRE, 17, Av.
des Martyrs, 38054 Grenoble Cedex 9, France}

\author{V. Leca}

\altaffiliation[Present address: ]{ Physikalisches Institut,
Universit\"{a}t T\"{u}bingen Auf der Morgenstelle 14, D-72076
T\"{u}bingen, Germany}

\author{D. H. A. Blank}

\author{H. Rogalla}

\author{H. Hilgenkamp}

\affiliation{Faculty of Science and Technology and MESA$^+$
Research Institute, University of Twente, P.O. Box 217, 7500 AE
Enschede, The Netherlands}

\date{\today}

\begin{abstract}
Phase-sensitive order parameter symmetry test experiments are
presented on the electron-doped high-$T_c$ cuprate {\nccoy}. These
experiments have been conducted using zigzag-shaped thin film
Josephson structures, in which the {\nccoy} is connected to the
low-$T_c$ superconductor Nb via a Au barrier layer. For the
optimally doped as well as for the overdoped {\nccoy} a clear
predominant \dwave-wave behavior is observed at {\tm}. Both
compounds were also investigated at $T=1.6$~K, presenting no
indications for a change to a predominant $s$-wave symmetry with
decreasing temperature.
\end{abstract}



\maketitle

The determination of the order parameter symmetry in the high
temperature superconductors is an important step towards the
identification of the mechanism of superconductivity in these
materials. This includes its dependencies on the sign and density
of the mobile charge carriers, on temperature and possible other
parameters. For the hole-doped high temperature superconductors,
such as {\ybco}, a long-lasting debate on the order parameter
symmetry was settled by the clear \dwave-wave behavior displayed
in various phase-sensitive symmetry test experiments, as reviewed
in~\cite{vanharlingen,Tsueirmp}. For the electron-doped materials,
Ln$_{2-x}$Ce$_x$CuO$_{4-y}$, with Ln = La, Nd, Pr, Eu or Sm,
$y\approx{0.04}$, only a few phase-sensitive test experiments have
until now been reported, all based on grain boundary Josephson
junctions. Tsuei and Kirtley~\cite{Tsuei} described the
spontaneous generation of half-integer flux quanta in {\nccoopy}
and Pr$_{1.85}$Ce$_{0.15}$CuO$_{4-y}$ tricrystalline films at
temperature {\tm}, presenting evidence for a \dwave-wave order
parameter symmetry. A similar conclusion was drawn by
Chesca~{\etal}~\cite{Chesca} from the magnetic field dependence of
the critical current for grain boundary-based $\pi$-SQUIDs in near
optimally doped {\lccoy}, also at {\tm}.

In contrast to these phase-sensitive experiments, a substantial
volume of more indirect symmetry test experiments exists for the
electron-doped materials. The conclusions from these studies are
varying. Behavior in line with an $s$-wave, or more general a
nodeless, symmetry was reported \eg, from the absence of a
zero-bias conductance peak in {\nccoopy} tunneling spectra at
$T\geq4.0$~K~\cite{Ekin,Kashiwaya,Alffprb} and from the
temperature dependencies of the London penetration depth in
Pr$_{1.855}$Ce$_{0.145}$CuO$_{4-y}$ for
1.6~K~$<T<$~24~K~\cite{Skintadtos}, in {\pccoy} with varying
Ce-content ($0.115\leq{x}\leq0.152$) for
0.5~K~$<T$~\cite{Kim087001}, and in {\nccoopy} for
1.5~K~$<T<$~4~K~\cite{Alff2644}, in addition to several earlier
studies~\cite{Anlage,Wu,Andreone}. On the other hand, $d$-wave
like characteristics were reported \eg, from the observed
gap-anisotropy in angle resolved photoemission spectroscopy on
{\nccoopy} at $T=10$~K~\cite{Armitage,Sato}, the temperature
dependence of the London penetration depth in optimally doped
{\pccoy} and {\nccoy} (0.4~K~$<T$)~\cite{Kokales,Prozorov} and
from the observation of zero-bias conductance peaks in optimally
doped {\nccoopy} ({\tm})~\cite{Hayashi} and
La$_{1.855}$Ce$_{0.105}$CuO$_{4-y}$ for
4.2~K~$<T<$~29~K~\cite{Chesca_Condmat0402131}.

Recently, a transition from $d$-wave behavior for underdoped
materials to $s$-wave like behavior for the optimally doped and
overdoped compounds was reported from the temperature dependence
of the London penetration depth in {\pccoy} and
{\lccoy}~\cite{Skintadtos} and from point contact
spectroscopy~\cite{Biswas,Qazilbash}. Further,
Balci~{\etal}~\cite{Balci} suggested a temperature-dependent
change in the order parameter symmetry for optimally and overdoped
{\pccoy}, with $s$-wave behavior at $T=2$~K and $d$-wave behavior
at $T\geq{3}$~K, based on specific heat measurements.

In view of this still ongoing discussion, there is a need for
further phase-sensitive experiments, and specifically to study
possible changes with temperature and doping. Tsuei and
Kirtley~\cite{Tsuei} and Chesca~{\etal}~\cite{Chesca} succeeded in
performing the first phase-sensitive measurements on the
electron-doped compounds based on grain boundary junctions.
Geometrical restrictions and the intrinsically low critical
current densities $J_c$ of the grain boundaries make such
experiments very challenging, especially for investigations on
non-optimally doped compounds. It is therefore advantageous to
also explore other Josephson junction configurations, with
potentially higher $J_c$'s. In addition, it would be very fruitful
to have a configuration for the symmetry test-experiment in which
a large Josephson penetration depth, associated with a low $J_c$,
presents an advantage rather than a difficulty. Both aspects are
fulfilled in the experiment described in the following, based on
zigzag-shaped Josephson contacts between {\nccoy} and Nb,
separated by a Au barrier layer.

\begin{figure}
\includegraphics[width=2.6in]{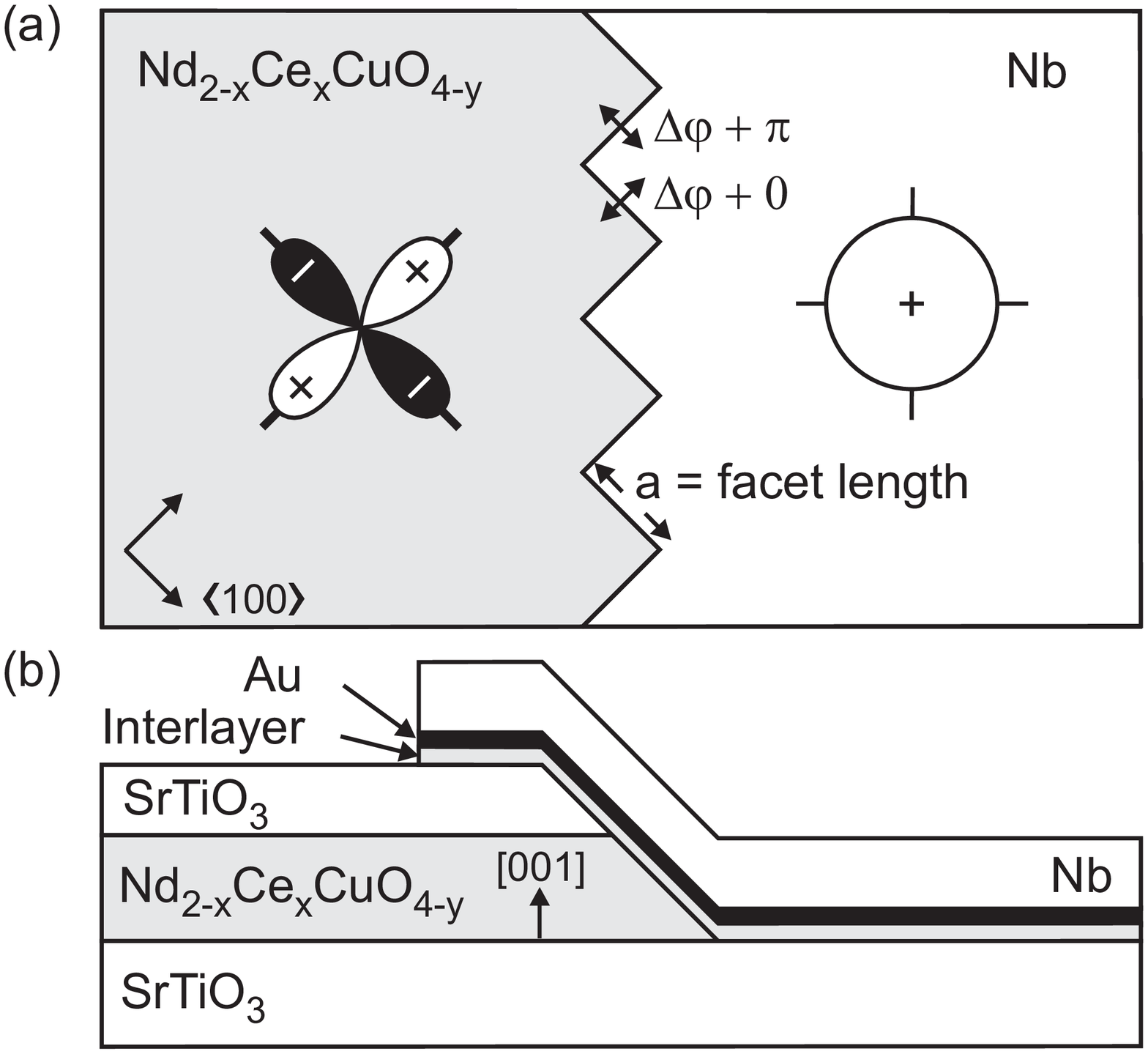}
\caption{\label{fig:ariando1}(a) Schematic topview of a \nccoy/Nb
zigzag structure with facet-length $a$. (b) Schematic sideview
illustrating the ramp-type \nccoy/Nb Josephson junction.}
\end{figure}

The zigzag-configuration (Fig.\ \ref{fig:ariando1}(a)) has been
described in detail in~\cite{Smildeprl}, where it was used to
investigate symmetry admixtures in \ybco. In these structures, all
interfaces are aligned along one of the
$\langle100\rangle$-directions of the cuprate, and are designed to
have identical $J_c$-values. With the high-$T_c$ cuprate being an
$s$-wave superconductor, the zigzag-structure presents no
significant difference to the case of a straight junction aligned
along one of the facet's directions. With the high-$T_c$
superconductor having a \dwave-wave symmetry, the facets oriented
in one direction experience an additional $\pi$-phase difference
compared to those oriented in the other direction. For a given
number of facets, the characteristics of these zigzag structures
then depend on the ratio of the facet length $a$ and the Josephson
penetration depth $\lambda_J$, see {\eg}~\cite{Zenchuk}. In the
small facet limit, $a\ll\lambda_J$, the zigzag structure can be
envisaged as a one-dimensional array of Josephson contacts with an
alternating sign of $J_c$, leading to anomalous magnetic field
dependencies of the critical current, as displayed for {\ybco} in
Ref.~\cite{Smildeprl}. In the large facet limit, the energetic
groundstate includes the spontaneous formation of half-integer
magnetic flux quanta at the corners of the zigzag structures, as
seen in~\cite{Hilgenkampnature}. All experiments on {\nccoy}
described below are in the small facet limit.

Figure~\ref{fig:ariando1}(b) schematically shows the \nccoy/Nb
ramp-type junctions that were used for the experiments. They were
prepared by first depositing a bilayer of 150~nm [001]-oriented
{\nccoy} and 35~nm SrTiO$_3$ by pulsed laser deposition on a
[001]-oriented SrTiO$_3$ single-crystal substrate. For the
optimally doped films a {\nccoop} target is used, for the
overdoped case a {\nccoov} target. The {\nccoy} film is grown at
820~$^\circ$C in 0.25~mbar of oxygen. Subsequently, the
temperature is reduced to 740~$^\circ$C for the growth of the
SrTiO$_3$ layer in 0.10~mbar of oxygen. Then the deposition
chamber is evacuated to about $10^{-6}$~mbar and the sample is
kept at 740~$^{\circ}$C for $10\--15$~minutes, before it is slowly
cooled down to room-temperature under vacuum conditions. For the
{\nccoopy} films, this procedure yields a typical critical
temperature $T_c$ of 20~K, the {\nccoovy} films had $T_c$'s of
13~K. The $T_c$'s for {\nccoopy} and {\nccoovy} were optimized
with respect to the oxygen content. The next step in the junction
fabrication process is the etching of a shallow ramp
($15\--20^{\circ}$) and cleaning of the sample using argon-ion
milling, analogous to the procedure used for
{\ybco}~\cite{Smildeprl,Smildeapl}. Then, using the same
deposition and cool-down conditions as for the first {\nccoy}
layer, a 12-nm {\nccoy} interlayer is deposited, followed by the
in-situ pulsed-laser deposition of a 12-nm Au barrier layer at
100~$^{\circ}$C. The interlayer, with the same composition as the
base-layer, is employed to provide an in-situ formed interface
between the cuprate layer and the Au-barrier. This is found to be
of great importance in reaching high junction quality, as it was
for the {\ybco} case~\cite{Smildeprl,Smildeapl}. Subsequently, a
160-nm Nb layer forming the counter electrode is deposited by
dc-sputtering and structured by lift-off. Finally, the redundant,
uncovered Au is removed by ion milling. In addition to zigzag
structures with different size and number of facets, every chip
contained several straight reference junctions oriented parallel
to one of the facet directions.

The junctions were characterized by measuring the current-voltage
($IV$) characteristics and the dependencies of the critical
currents $I_c$ on applied magnetic field $H_a$, using a four-probe
method with the magnetic field parallel to the [001]-direction of
the {\nccoy} in a well-shielded cryostat at {\tm} and $T=1.6$~K.
For the determination of $I_c$, a typical voltage criterion of
$V_c\lesssim2~\mu$V was used. This yields a lower limit of
$V_c/R_n$ to $I_c$, with $R_n$ being the junctions' normal state
resistance. In the $I_c(H_a)$-dependencies presented here, the
lowest $I_c$ values can be considered as the lower limit in the
determination of the critical current for that respective
measurement.

\begin{figure}
\includegraphics[width=2.6in]{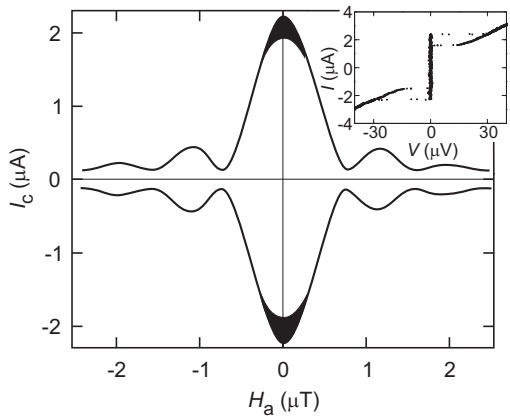}
\caption{\label{fig:ariando2}Critical current $I_c$ as a function
of applied magnetic field $H_a$ for a 50~$\mu$m wide straight
\nccoop/Nb ramp-type junction ({\tm}). The dark areas correspond
to the hysteresis in the current-voltage characteristic shown for
$H_a=0$ in the inset.}
\end{figure}

Figure \ref{fig:ariando2} shows the {\icb}-dependence recorded for
a 50~$\mu$m wide straight \nccoopy/Nb reference junction at {\tm},
and in the inset its zero-field $IV$-characteristic. The
\icb-dependence closely resembles a Fraunhofer pattern, which is
the hallmark of small rectangular junctions with homogeneous
current distributions. A maximum $I_c=2.2~\mu$A at zero applied
field was found. The black areas in the peaks of the {\icb} curves
are indicative for the hysteresis in the $IV$-characteristics. At
the measuring temperature, this junction has a typical $J_c$ of
${29}$~A/cm$^2$, from which a value for the Josephson penetration
depth $\lambda_J=65~\mu$m is estimated. This $J_c$ is several
times larger than attainable with grain boundary junctions. The
normal-state resistance for this junction is 13~$\Omega$, which
gives an $I_cR_n$ product of about 30~$\mu$V and
$R_nA=1.0\times10^{-6}~\Omega$cm$^2$.

\begin{figure}
\includegraphics[width=2.6in]{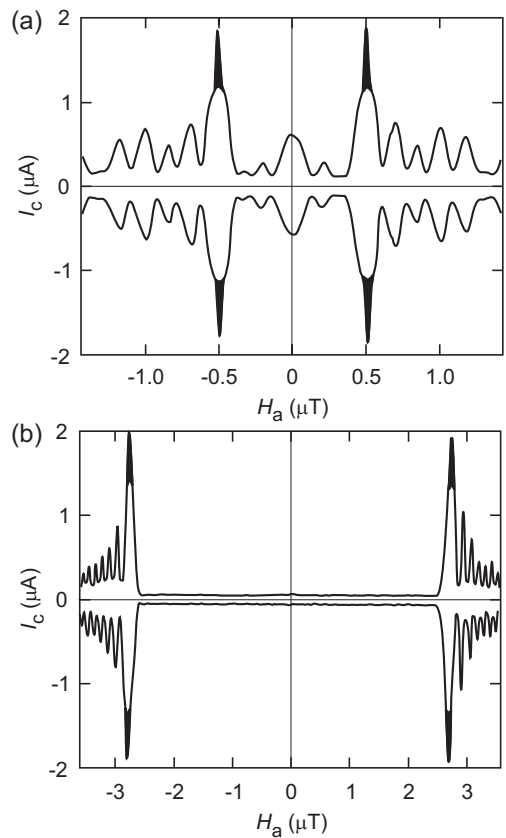}
\caption{\label{fig:ariando3}Critical current $I_c$ as a function
of applied magnetic field $H_a$ for (a) a \nccoop/Nb zigzag array
comprised of 8 facets of 25~$\mu$m width and (b) a similar array
with 80 facets of {5 $\mu$m} width ({\tm}).}
\end{figure}

The \icb-dependence for a \nccoop/Nb zigzag junction having 8
facets of 25~$\mu$m width is presented in
Fig.~\ref{fig:ariando3}(a). Instead of an $I_c$-maximum at
$H_a=0$, one can observe a maximum $I_c$ of 1.8~$\mu$A at
$H_a=0.5~\mu$T. This zigzag junction shows a highly symmetric
{\icb} pattern for both polarities of current bias and applied
magnetic field. The critical current at $H_a=0$ falls to less than
32\% of its peak value. Presuming that $J_c$ for this junction is
equal to the reference junction described above, the zero field
$I_c$ is only 7\% of the expected value for an equally long
straight junction, disregarding wide-junction effects. It should
be noted that also the maximum $I_c$ at $H_a=0.5~\mu$T is $2\--3$
times lower than expected based on the $J_c$ of the straight
junction. Small variations in the thickness of the Au
barrier-layer or the {\nccoop} interlayer between the straight
junction and the zigzag structure, placed several mm from each
other on the chip, may well account for a considerable part of
this $J_c$-difference. Further, as was shown by Zenchuk and
Goldobin~\cite{Zenchuk}, a zigzag structure with an odd number of
corners is expected to produce spontaneous magnetic flux for all
facet-lengths. As the \icb-dependence of
Fig.~\ref{fig:ariando3}(a) is still strongly non-Fraunhoferlike,
this spontaneous flux is expected to be smaller than a half-flux
quantum per facet length. Nevertheless, a part of the $I_c$
observed at $H_a=0$ and the reduced peak height at $H_a=0.5~\mu$T
may be resulting from this spontaneous flux.

In Fig.\ \ref{fig:ariando3}(b), the {\icb}-dependence for a zigzag
array with 80 facets having a substantially smaller facet length
of 5~$\mu$m is shown, presenting a maximum $I_c=2.0~\mu$A at
$H_a=2.8~\mu$T. Also for this very dense zigzag structure the
{\icb}-dependence is highly symmetric. For this structure, a very
low ratio of 2\% between the critical current at zero magnetic
field and the maximal critical current is found.

The \icb-dependencies of these zigzag structures clearly exhibit
the characteristic features also seen for the {\ybco}
case~\cite{Smildeprl}, namely the absence of a global maximum at
$H_a=0$ and the sharp increase in the critical current at a given
applied magnetic field. This behavior can only be explained by the
facets being alternatingly biased with or without an additional
$\pi$-phase change. This provides a direct evidence for a
$\pi$-phase shift in the pair wave function for orthogonal
directions in momentum space and thus for a predominant {\dwave}
order parameter symmetry.

If the order parameter were to comprise an imaginary $s$-wave
admixture, the {\icb}-dependencies for the zigzag junctions would
be expected to display distinct asymmetries, especially for low
fields~\cite{Smildeprl}. In addition, the critical current at zero
applied field is expected to increase with the fraction of
$s$-wave admixture. From the high degree of symmetry of the
measured characteristics of Figs~\ref{fig:ariando3}(a) and
\ref{fig:ariando3}(b) and the very low zero field $I_c$, no sign
of an imaginary $s$-wave symmetry admixture to the predominant
{\dwave} symmetry can be distinguished.

\begin{figure}
\includegraphics[width=2.8in]{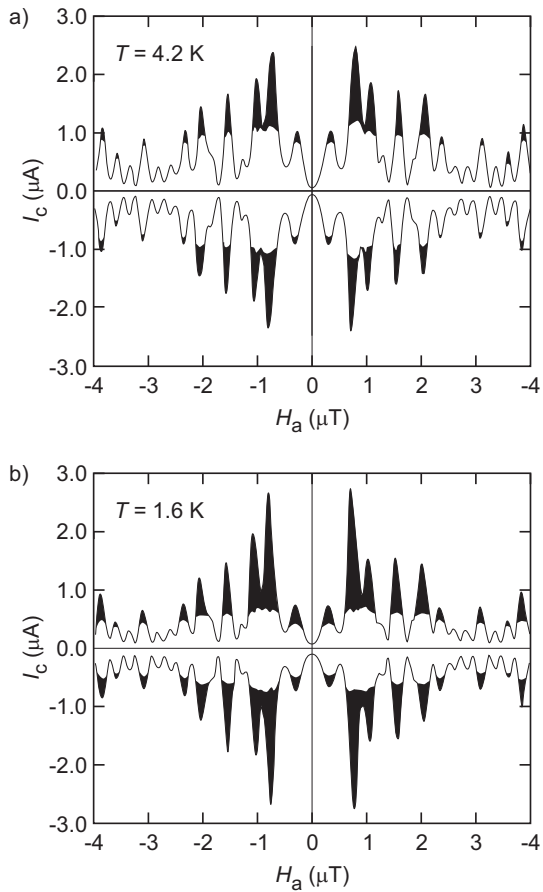}
\caption{ \label{fig:ariando4}Critical current $I_c$ as a function
of applied magnetic field $H_a$ for a \nccoov/Nb zigzag array
comprised of 8 facets of 25~$\mu$m width ({\tm}) at (a)~$T=4.2$~K
and (b)~$T=1.6$~K.}
\end{figure}

To investigate a possible change of the order parameter symmetry
with doping we have fabricated similar zigzag structures using
\nccoov/Nb junctions. Figure~\ref{fig:ariando4}(a) shows the
{\icb}-dependence measured at {\tm} for a structure with 8 facets
of 25~$\mu$m width. Obviously, also these characteristics indicate
a predominant \dwave-wave symmetry.

When cooling down the samples to $T=1.6$~K all the basic features
displayed by the structures at {\tm} remain unaltered, as is shown
for the overdoped sample in Fig.~\ref{fig:ariando4}(b). We thus
see no indication for an order parameter symmetry crossover for
{\nccoopy} in this temperature range, as was recently reported for
{\pccoy}~\cite{Balci}. Similar results were obtained for optimally
doped samples upon cooling down to $T=1.6$~K.

In conclusion, our phase-sensitive order parameter symmetry test
experiments based on {\nccoy}-Nb zigzag junctions provide clear
evidence for a predominant {\dwave} order parameter symmetry in
the {\nccoy}. This corroborates the conclusions of studies
performed with grain boundary junctions in the optimally doped
compounds. To verify various recent reports on possible order
parameter changes with overdoping and with decreasing temperature,
we have studied the influence of those parameters. No change in
the symmetry was observed when overdoping the {\nccoy} compound.
Further, the order parameter symmetry was found to remain
unaltered between $T=1.6$~K and {\tm}. This study does not provide
an explanation for the contradicting results obtained in other
experiments.

\begin{acknowledgments}
The authors thank A. Brinkman, S. Harkema and G. Rijnders for
helpful discussions. This work was supported by the Dutch
Foundation for Research on Matter (FOM), the Netherlands
Organization for Scientific Research (NWO) and the European
Science Foundation (ESF) PiShift programme.
\end{acknowledgments}


\begin{thebibliography}{99}

\bibitem{vanharlingen} D. J. Van Harlingen, Rev. Mod. Phys. {\bf 67}, 515 (1995).
\bibitem{Tsueirmp} C. C. Tsuei and J. R. Kirtley, Rev. Mod. Phys. {\bf 72}, 969 (2000).
\bibitem{Tsuei} C. C. Tsuei and J. R. Kirtley, Phys. Rev. Lett. {\bf 85}, 182 (2000).
\bibitem{Chesca} B. Chesca \etal, Phys. Rev. Lett. {\bf 90}, 057004 (2003).
\bibitem{Ekin} J. W. Ekin \etal, Phys. Rev. B {\bf 56}, 13746 (1997).
\bibitem{Kashiwaya} S. Kashiwaya \etal, Phys. Rev. B {\bf 57}, 8680 (1998).
\bibitem{Alffprb} L. Alff \etal, Phys. Rev. B {\bf 58}, 11197 (1998).
\bibitem{Skintadtos} J. A. Skinta \etal, Phys. Rev. Lett. {\bf 88}, 207005 (2002).
\bibitem{Kim087001} M. -S. Kim \etal, Phys. Rev. Lett. {\bf 91}, 087001 (2003).
\bibitem{Alff2644} L. Alff \etal, Phys. Rev. Lett. {\bf 83}, 2644 (1999).
\bibitem{Anlage} S. M. Anlage \etal, Phys. Rev. B {\bf 50}, 523 (1994).
\bibitem{Wu} D. H. Wu \etal, Phys. Rev. Lett. {\bf 70}, 85 (1993).
\bibitem{Andreone} A. Andreone \etal, Phys. Rev. B {\bf 49}, 6392 (1994).
\bibitem{Armitage} N. P. Armitage \etal, Phys. Rev. Lett. {\bf 86}, 1126 (2001).
\bibitem{Sato} T. Sato \etal, Science {\bf 291}, 1517 (2001).
\bibitem{Kokales} J. D. Kokales \etal, Phys. Rev. Lett. {\bf 85}, 3696 (2000).
\bibitem{Prozorov} R. Prozorov \etal, Phys. Rev. Lett. {\bf 85}, 3700 (2000).
\bibitem{Hayashi} F. Hayashi \etal, J. Phys. Soc. Jpn. {\bf 67}, 3234 (1998).
\bibitem{Chesca_Condmat0402131} B. Chesca \etal, Condmat-0402131(2003).
\bibitem{Biswas} A. Biswas \etal, Phys. Rev. Lett. {\bf 88}, 207004 (2002).
\bibitem{Qazilbash} M. M. Qazilbash \etal, Phys. Rev. B {\bf 68}, 024502 (2003).
\bibitem{Balci} H. Balci and R. L. Greene, Cond-mat 0402263 (2004).
\bibitem{Smildeprl} H. J. H. Smilde \etal, Phys. Rev. Lett. {\bf 88}, 057004 (2002).
\bibitem{Zenchuk} A. Zenchuk and E. Goldobin, Phys. Rev. B {\bf 69}, 024515 (2004).
\bibitem{Hilgenkampnature} H. Hilgenkamp \etal, Nature {\bf 422}, 50 (2003).
\bibitem{Smildeapl}H. J. H. Smilde \etal, Appl. Phys. Lett. {\bf 76}, 912 (2002).

\end{thebibliography}

\end{document}